\documentclass[11pt]{article}
\usepackage{jheppub}
\usepackage{amsmath,amssymb,amsfonts,graphicx,slashed,color, mathtools, amsthm}
\usepackage{comment}
\usepackage{enumerate}
\usepackage{float}
\usepackage{caption}
\usepackage{subcaption}
\usepackage{pdflscape}

\newcommand{\be}{\begin{equation}}
\newcommand{\ee}{\end{equation}}
\newcommand{\bea}{\begin{eqnarray}}
\newcommand{\eea}{\end{eqnarray}}
\newcommand{\beas}{\begin{eqnarray*}}
\newcommand{\eeas}{\end{eqnarray*}}
\newcommand{\ba}{\begin{array}}
\newcommand{\ea}{\end{array}}

\usepackage{xcolor}      

\title{Cosmology with non-conformal holographic matter}

\author{Mark Van Raamsdonk,}
\author{Rana Zibakhsh}

\affiliation{Department of Physics and Astronomy, University of British Columbia,\\
6224 Agricultural Road, Vancouver, B.C.\ V6T 1Z1, Canada}

\emailAdd{mav@phas.ubc.ca}
\emailAdd{rzibakhsh@phas.ubc.ca}

\abstract{We investigate the effect on cosmological evolution of a strongly coupled quantum field that undergoes renormalization group flow from a UV CFT to an IR CFT. The field theory is defined by perturbation of a holographic CFT by a relevant operator associated with a bulk scalar field that evolves from a local maximum of its potential near the boundary to a local minimum of its potential deep in the bulk. By studying the gravity solutions dual to this theory on $\mathbb{R}^3 \times S^1$, we find that the equation of state parameter $w$ for the field theory has the conformal behavior $w=1/3$ for high and low temperatures, but dips to lower values for intermediate temperatures. Thus, at scales where the field theory has significant scale-dependence, its effect on cosmological evolution is intermediate between matter and radiation. Compared to the unperturbed UV CFT (which acts as radiation), the energy density experiences less dilution during the expansion as a result of the RG flow, and the rate of expansion is greater.
}

\begin{document}

\maketitle

\section{Introduction}

Ordinary matter described by the Standard Model is believed to comprise only a small fraction of the total energy density in the universe. The remaining components, dubbed dark matter and dark energy, are still mysterious \cite{hawking_ellis_1973, weinberg_cosmology_2008, Frieman2008, 1940oort}. In the $\Lambda$CDM model of cosmology, the dark matter is assumed to be pressureless matter with equation of state parameter $w = 0$, while the dark energy is taken to be a cosmological constant with $w = -1$ \cite{Bahcall_1999}.\footnote{The equation of state parameter is defined as $w = p/e$ where $p$ is the pressure and $e$ is the energy density. In general, this can depend on the temperature.} But the dark sector might well be more complicated or include additional components. It is common in observational studies to consider phenomenological models with different, possibly time-dependent equation of state parameters and to constrain the possible behaviors of these \cite{darkenergy,dark_energy}.

An interesting possibility for the dark sector is that it contains a component described by some strongly interacting quantum field theory (see, for example \cite{Spergel2000,Kaplinghat2016,Feng_2021,Fitzpatrick_2013, dark_holog, dynamicalinflaton}). If this field theory is conformal, the equation of state is the same as for radiation. But for a non-conformal field theory with non-trivial RG flow, the behavior could be more interesting. Our goal in this paper is to understand how the equation of state parameter behaves as a function of temperature for such a theory and understand the effects on cosmological evolution.

In order to make the analysis tractable, we consider a quantum field theory that can be studied using holography (the AdS/CFT correspondence) \cite{maldacena_large_1999,aharony_large_2000}. We perturb a holographic CFT by some relevant operator associated with a bulk scalar field with $m^2 < 0$. With a potential $V(\phi) = 1/2 m^2 \phi^2 + c/3 \phi^4$, we have a solution where the scalar evolves from the local maximum at $\phi=0$ in the UV to the local minimum for $\phi > 0$ deep in the IR. The solution is locally AdS both in the UV and in the IR, with a larger magnitude cosmological constant in the IR, corresponding to the reduced number of degrees of freedom in the IR of the field theory. To study the thermodynamics, we study the theory on $\mathbb{R}^3 \times S^1$, constructing the dual gravity solutions numerically and analyzing them to extract the thermodynamic behavior of the field theory. For the thermodynamic analysis, we employ a technique that avoids having to introduce bulk counterterms and calculate the regularized action. Instead, we extract the temperature and entropy density from the solution (looking at the periodicity of the Euclidean time direction and the horizon area respectively) and find the pressure and energy density by integrating thermodynamic relations $de = Tds$ and $dp = s dT$.

Using the gravity calculations, we determine the behavior of the equation of state parameter $w$ as a function of temperature in the space of theories parameterized by $m^2$ (related to the dimension of the relevant operator we are perturbing by) and $c$ (which controls the endpoint of the RG flow). In each case, we find that $w$ approaches the conformal value $w=1/3$ for high and low temperatures, consistent with the fact that these regimes are governed by the UV and the IR CFT. For intermediate temperatures, we find that the equation of state parameter dips to smaller values. This decrease in $w$ is more pronounced for smaller $m^2$ (lower dimension operators) and for smaller values of $c$, both of which lead to a lower minimum of the potential and thus a greater reduction in the number of degrees of freedom along the RG flow.

We then couple the holographic field theory to (4D) gravity and study the cosmological evolution in a quasi-static approximation where we assume the equilibrium thermodynamics remains valid during the evolution.\footnote{We check that this approximation is valid for the parameter values we consider by ensuring that the equation of state parameter changes slowly on the Hubble timescale.} If we take the holographic field theory to be the only matter in the cosmology, there is a unique flat cosmological solution for each choice of the parameters $m^2$ and $c$. This has the $a \propto t^{1/2}$ expansion characteristic of a CFT for early and late times but an intermediate regime where the expansion speeds up relative to theory where the matter is taken to be the unperturbed UV CFT. 

In the body of the paper below, we describe the field theory setup in section 2, explain the details of the holographic calculations in section 3, present the detailed results in section 4, and end with a brief discussion in section 5.

\subsubsection*{Relation to previous work}

There have been many other works in the past investigating the thermodynamic behavior of holographic field theories or aspects of their cosmological physics. Papers that specifically consider aspects of equilibrium and non-equilibrium thermodynamics for non-conformal holographic theories include \cite{Attems_2016,G_rsoy_2018, Kleinertt_2016, Attems_2017, Ballon_QCD_2021,Buchel_QCD_2016,G_rsoy_QCD_2016}.
Various papers have studied the behavior of holographic field theories on fixed cosmological backgrounds, including \cite{Ghosh_2018, Casalderrey_Solana_2021, Marolf_2011}. 
Other papers that have studied cosmological evolution of holographic field theories dynamically coupled to gravity include \cite{2022paper, dynamicalinflaton, Koyama2001} where the authors go beyond the quasi-static approximation to study situations where the quantum field theory may be out of equilibrium. Holographic non-conformal matter has also been studied previously in the context of braneworld cosmology, with equation-of-state parameters arising from scalar potentials different from those considered in this paper (for related work, see \cite{Fichet:2022ixi,Fichet:2022xol,Fichet:2023dju}).
In \cite{reheating,reheating2, reheating3, warminflation} the authors consider holographic field theories as part of the physics of inflation; an important aspect of strongly coupled QFTs is that they quickly thermalize and enter the hydyodynamic regime making them potentially useful for facilitating the exit from inflation to the hot Big Bang. 

Novel aspects of this work include  
i) a systematic study of how the scale dependent equation of state parameter varies with the scaling dimension of the perturbing operator and with the length of the RG flow, using perhaps the simplest holographic QFT that interpolates between UV and IR CFTs;
ii) the use of a numerical method that obtains thermodynamic quantities directly from bulk solutions without introducing explicit holographic counterterms.  While this technique has appeared before in other contexts (see, e.g.\ \cite{G_rsoy_2009}), here we adapt it to track cosmological solutions along the RG flow and to map out the full $w(\mathcal{T})$ curve; and
iii) a numerical exploration of the dynamical evolution of the scale factor in spatially flat cosmologies where the only matter component is the strongly coupled non-conformal holographic theory.

\section{Quantum field theory setup}

In this paper, we consider non-conformal quantum field theories that have an RG flow from a UV CFT to an IR CFT. We would like to understand how such a field theory contributes to background cosmological evolution, for example if it is present as a part of the dark sector in our universe. In the next section, we will specialize to the case of a holographic field theory, but for now, we will not assume this.

\subsection{Equilibrium thermodynamics in Minkowski space}

To begin, we consider thermal physics of our quantum field theory on a $3+1$ dimensional Minkowski background. The theory is defined via deformation of a conformal field theory by a relevant operator ${\cal O}$ of dimension $\Delta < 4$. We can write the action as
\begin{equation}
    S_1 = \textit{S}_{0} + \int d^4 x \textit{J} \ {\cal O}(x) \; ,
\end{equation}
where $\textit{S}_{0}$ is the CFT action and $\textit{J}$ is a spacetime-independent source. In order to study the thermodynamics, we consider the Euclidean theory on $\mathbb{R}^3 \times S^1$, where the thermal circle has length $\beta = 1/T$. From dimensional analysis, $[J]=4-\Delta$, so we can define a dimensionless temperature parameter
\begin{equation}
    \mathcal{T} \equiv T J^{1/(\Delta - 4)} \; .
\end{equation}
We would like to understand the behavior of the equilibrium entropy density $s$, energy density $e$ and pressure $p$ as a function of this parameter. These allow us to establish the equation of state that will feed into our subsequent cosmological analysis.\footnote{We will discuss below the conditions under which it is a good approximation to use equilibrium thermodynamics.} In particular, we will be interested in the behavior of the equation of state parameter $w = p/e$ as a function of $\mathcal{T}$. 

In the holographic field theories that we consider below, it will be simplest to read off the entropy density as a function of temperature from the gravity solutions, so it will be convenient to express the other thermodynamic quantities in terms of this. 
Starting from the First Law of Thermodynamics $dE = T dS - p dV$ and the definition of free energy $F = E - T S$, we have that $dF = -S dT - p dV$, so that 
\begin{equation}
    p = - \left. {\frac{\partial F}{\partial V}} \right|_T
\end{equation} 
For the systems we consider, the free energy is extensive, so the pressure is simply equal to the negative of the free energy density. Then working at a fixed volume (that we can take to infinity), we have
\begin{equation}
\label{tsdiff}
     e + p = T s  \qquad \qquad de = T ds \qquad \qquad dp = s dT \; .
\end{equation} 

It will be useful below to work with dimensionless versions of the entropy density $s$, energy density $e$ and pressure $p$. These have dimensions 3, 4, and 4 respectively, so we can write
\begin{equation}
    s = T^3 {\cal S}(\mathcal{T}), \qquad e = T^4 {\cal E}(\mathcal{T}) ,\qquad  p = T^4 {\cal P}(\mathcal{T})
\end{equation}
for some dimensionless functions ${\mathcal{S}} (\mathcal{T}), {\mathcal{E}}(\mathcal{T})$ and ${\cal P}(\mathcal{T})$. From (\ref{tsdiff}) we find that these satisfy 
\begin{equation}
\label{dimleqns}
    {\cal E} + {\cal P} = {\cal S} \qquad 4 {\cal E} + {\cal T} {d {\cal E} \over d {\cal T}} = 3 {\cal S} + {\cal T} {d {\cal S} \over d {\cal T}} \qquad 4 {\cal P} + {\cal T} {d {\cal P} \over d {\cal T}} =  {\cal S} 
\end{equation}
Each of these functions is a constant for a CFT, and the thermodynamic relations above imply that these constants are related by
\begin{equation}
\label{CFTrels}
    {\cal E}_{CFT} = {3 \over 4} {\cal S}_{CFT} \qquad \qquad {\cal P}_{CFT} = {1 \over 4} {\cal S}_{CFT} \; .
\end{equation}
For our RG flow theory, ${\cal S}$ should approach constants ${\cal S}_{UV}$ and ${\cal S}_{IR}$ for large and small $\mathcal{T}$ with the dimensionless energy and pressure related to these as in (\ref{CFTrels}).

The relations (\ref{tsdiff}) can be expressed in terms of these dimensionless quantities and integrated to give
\begin{equation}
    \mathcal{P}(\mathcal{T}) = 1/\mathcal{T}^4 \int_{0}^{\mathcal{T}}  \tilde{\mathcal{T}}^3 \mathcal{S} (\tilde{\mathcal{T}}) d\tilde{\mathcal{T}} \qquad \qquad \mathcal{E} (\mathcal{T}) = \mathcal{S}(\mathcal{T}) - \mathcal{P}(\mathcal{T})
\end{equation}
We will use these relations below to calculate $\mathcal{E}$, $\mathcal{P}$, and $w = \mathcal{P}/\mathcal{E}$ for our holographic field theory starting from $\mathcal{S}(\mathcal{T})$ which can be read off the gravity solutions.

\subsection{Cosmology}

Now, suppose that we have such a field theory coupled to gravity and consider the cosmological solutions that arise when this field theory provides the only source of energy and pressure.\footnote{We will discuss the case with additional sources of energy below.} In this case, assuming a flat cosmology, we have metric
\begin{equation}
    ds^2 = -dt^2 + a^2(t) d\vec{x}^2 \; .
\end{equation}
Defining the Hubble parameter $H = \dot{a}/a$, the evolution of the scale factor is determined by the Friedmann equation 
\begin{equation}
H^2 = \frac{8 \pi G_4}{3} e
\end{equation}
together with the continuity equation
\begin{equation}
    \frac{de}{dt} = -3H(e+p),
\end{equation}
and the equation of state relating $e$ and $p$. Using (\ref{tsdiff}), we can rewrite this as
\begin{equation}
    {ds \over s} = -3 {da \over a}
\end{equation}
This gives
\begin{equation}
    s = {s_0 \over a^3}
\end{equation}
where $s_0$ is the entropy density when $a=1$. Of course, this is just the statement that the evolution represents an adiabatic expansion, a consequence of our assumption that equilibrium thermodynamics is valid. We expect this to fail if the expansion is too fast; we discuss this more in section 4.1 below.
 
Translating into the dimensionless quantities, we can write the scale factor at a given dimensionless temperature as
\begin{equation}
\label{asol}
    a = {{\cal T}_1 \mathcal{S}^{1 \over 3}({\cal T}_1) \over {\cal T} \mathcal{S}^{1 \over 3}({\cal T})}
\end{equation}
We can set ${\cal T}_1$ to any convenient value since the scale factor can be rescaled using a coordinate transformation that rescales the spatial coordinates. Below, it will be convenient to choose ${\cal T}_1$ such that
\begin{equation}
\label{asol}
    a = { \mathcal{S}_{UV}^{1 \over 3} \over {\cal T} \mathcal{S}^{1 \over 3}({\cal T})}
\end{equation}

We can find the time as a function of the dimensionless temperature (and thus the scale factor as a function of time) using the Friedmann equation, which can be rewritten (assuming expansion) as
\begin{equation}
\label{fried2}
    \frac{da}{dt} =  \sqrt{\frac{8\pi G_4}{3}} \sqrt{e} a \; .
\end{equation}
Defining a dimensionless time parameter 
\begin{equation}
    \tau =  \sqrt{\frac{8\pi G_4}{3}} J^{2 \over 4 - \Delta} t
\end{equation}
we have that
\begin{equation}
\label{seq}
    {d\tau \over d {\cal T}} =  -\left( {1 \over {\cal T}} +{1 \over 3 {\cal S}} {d {\cal S} \over d \cal T}\right) {1 \over {\cal T}^2 \sqrt{{\cal S} - {\cal P}}}
\end{equation}

Given ${\cal S}({\cal T})$, this can be integrated together with the last equation in (\ref{dimleqns}) to determine $\tau({\cal T})$. Together with (\ref{asol}), this determines the scale factor evolution parametrically. If we define $\tau=0$ to be the time for which ${\cal T} \to \infty$ and normalize $a$ as above, there is a unique solution.

With additional matter components, we can determine the evolution by replacing $e \to e + \sum_i e_i(a)$ in the Friedman equation (\ref{fried2}) where $e_i(a)$ describe the energy density as a function of scale factor for these other components.

For conformal matter, recalling that ${\cal S}$, ${\cal E}$, and ${\cal P}$ are constant, we have from equation
(\ref{asol}) that 
\begin{equation}
    a = {{\cal T}_0 \over {\cal T}}
\end{equation}
while (\ref{seq}) simplifies to
\begin{equation}
\label{seqconf}
    {d\tau \over d {\cal T}} =  -{1 \over {\cal T}^3}  {1 \over \sqrt{{\cal E}_{CFT}}}
\end{equation}
giving
\begin{equation}
    \tau =  {1 \over 2} {1 \over {\cal T}^2 \sqrt{{\cal E}_{CFT}}}; .
\end{equation} 
Choosing ${\cal T}_0 =1$, we have 
\begin{equation}
    a(\tau) = \sqrt{2 \sqrt{{\cal E}_{CFT}}} \sqrt{\tau} \; .
\end{equation}
For this evolution, we have a constant deceleration parameter
\begin{equation}
    q \equiv - {\ddot{a} a \over \dot{a}^2} = -1 \; .
\end{equation}
For the RG flow theory, we expect this $a \propto \sqrt{\tau}$ behavior for small and large $\tau$ with ${\cal E}_{CFT}$ replaced with ${\cal E}_{UV}$ and ${\cal E}_{IR}$ respectively. The deceleration parameter will approach -1 for small and large times, but deviate from 1 at intermediate times. Using the Friedmann equation, along with the acceleration equation
\begin{equation}
    -{\ddot{a} \over a} = {4 \pi G \over 3} (3 p + e)
\end{equation}
we have
\begin{equation}
q = - {1 \over 2} (1 + 3 w)
\end{equation}
where $w = p/e = {\cal P} / {\cal E}$ so the deviations of scale factor deceleration from the CFT behavior $q=-1$ are directly related to the deviations of the equation of state parameter $w$ from the CFT behavior $w=1/3$. Calculating $w$ as a function of ${\cal T}$ for various holographic field theories will be one of our main goals below.

\section{Holographic calculations}

We now specialize to the case of a holographic quantum field theory defined by a dual 4+1 dimensional gravitational system which involves a metric minimally coupled to a classical scalar field. The Euclidean action is 
\begin{equation}
    S_{bulk} = \frac{1}{16 \pi G_5}\int dx^5 \sqrt{g} \big( R + \frac{12}{L_{AdS}^2} + (\partial \phi)^2 + 2 V(\phi) \big),
\end{equation}
where $G_5$ is the five-dimensional Newton's constant. For solutions corresponding to the field theory on $\mathbb{R}^3 \times S^1$, we consider a metric ansatz given by:
\begin{equation}
    ds^2 = f(r) e^{-\chi(r)}dt^2 + \frac{1}{f(r)} dr^2 + r^2 d\vec{x}^2,
    \label{metric}
\end{equation}
where $\vec{x} = (x,y,z)$ correspond to the spatial coordinates of the field theory, $t$ is the Euclidean time coordinate that we will take to be periodic, and $r$ is the radial coordinate (related to the energy/distance scale in the dual field theory). 

The RG-flow in the field theory is reflected in a radially-dependent scalar field
\begin{equation}
    \phi = \phi(r) \; .
    \label{metric}
\end{equation}
The choice of the potential $V(\phi)$ determines the properties of this RG flow. We consider a simple potential 
\begin{equation}
    V(\phi) = \frac{1}{2} m^2 \phi^2 + \frac{c}{3} \phi^4 
\end{equation}
with an extremum at $\phi=0$ corresponding to the UV fixed point and a minimum for some $\phi_{min} > 0$ corresponding to the IR fixed point. The mass parameter $m^2$ is related to the dimension of the dual scalar operator through $m^2 = \Delta (\Delta-4)$. For fixed $m^2$, varying the parameter $c$ changes the value of $\phi_0$ at the minimum and the potential value $V(\phi_{min})$ that sets the effective cosmological constant and AdS length in the IR region.

The solution must satisfy the Einstein equation $G_{\mu \nu} +\Lambda g_{\mu \nu} = T_{\mu \nu}$ where 
\begin{equation}
    T_{\mu \nu} = -\frac{1}{2} g_{\mu \nu} g^{\rho \sigma} \nabla_\rho \phi \nabla_\sigma \phi + \nabla_\mu \phi \nabla_\nu \phi - g_{\mu \nu} V(\phi),
\end{equation}
is the scalar field stress-energy tensor. We find that this is satisfied provided that
\begin{equation}
    \chi'+\frac{2r}{3} \phi'^2 =0
    \label{einstein1}
\end{equation}
and
\begin{equation}
    f' + (\frac{2}{r}-\frac{\chi'}{2})f -4r +\frac{2r}{3} V = 0,
    \label{einstein2}
\end{equation}
where we have set $L_{AdS}=1$, giving $\Lambda = -6$ for the $4+1$ dimensional asymptotically AdS background. 
We also have the scalar field equation of motion $- \nabla^2 \phi + V'(\phi) = 0$ which leads to
\begin{equation}
\label{KG}
    \phi'' + (\frac{3}{r} - \frac{\chi'}{2}+\frac{f'}{f})\phi' = \frac{1}{f} \frac{dV}{d\phi} \; .
\end{equation}

The equations eq. \ref{einstein1} and eq. \ref{einstein2} are invariant under the following transformations:
\begin{equation}
\label{confsym}
    \tilde{\phi}(r) = \phi(ar), \ \ \tilde{f}(r) = \frac{1}{a^2} f(ar), \ \ \tilde{\chi}(r) = \chi(ar) 
\end{equation}
and 
\begin{equation}
\label{rescale}
    \tilde{\chi} = \chi + C.
\end{equation}
These symmetries arise from the underlying conformal invariance of the dual CFT and the freedom to rescale the time coordinate.

The solutions that we will study approach the AdS metric asymptotically as $r \rightarrow \infty$ reflecting the fact that the UV fixed point is a holographic CFT. This asymptotic behavior corresponds to 
\begin{equation}
    f(r) \to r^2 \qquad \chi(r) \to 0 \qquad \phi(r) \to 0 \qquad \qquad r \to \infty \; .
\end{equation}
Via the standard AdS/CFT dictionary, the source $J$ in for the operator associated with the scalar can be read off from the asymptotic behavior of the scalar field. Specifically, $J$ is the coefficient of the term proportional to $r^{\Delta -4}$ in the asymptotic expansion, which is the leading term when $\Delta > 2$. In this case
\begin{equation}
\label{Jasympt}
    J = \lim_{r \to \infty} r^{4 - \Delta} \phi(r).
\end{equation}

For some $r = r_H$ that corresponds to a horizon in the Lorentzian solution, we have that $f \to 0$. To avoid a conical singularity in the Euclidean solution, the periodicity $\beta = 1/T$ of the Euclidean time must satisfy 
\begin{equation}
    T = \frac{f'(r_H) e^{-\chi(r_H)/2}}{4\pi} \; .
\end{equation}

We can parameterize the physically distinct solutions by the value of the scalar field $\phi_0 = \phi(r_H)$ at this horizon point. The derivative $\phi'(r)$ is fixed in terms of $\phi(r)$;  multiplying the (\ref{KG}) by $f$ and evaluating at $r_H$, and combining this with (\ref{einstein2}) evaluated at $r_H$, we obtain
\begin{equation}
    \phi'(r_H) = \frac{3dV/d\phi|_{r_H}}{2r_H(6 -V(r=r_H))} \; .
\end{equation}
This equation, together with the remaining boundary conditions
\begin{equation}
    f(r_H) =0, \ \ \phi(r_H) = \phi_0, \  \ \chi(\infty) = 0, \ \
\end{equation}
fixes a solution of our three equations. In practice, we can take $\chi(r_H) = 0$ as a boundary condition when solving the equations, and then use the symmetry (\ref{rescale}) to subtract $\chi(\infty)$ from $\chi(r)$ to enforce the boundary condition at infinity. 

The remaining symmetry (\ref{confsym}), corresponding to the scaling symmetry in the field theory, relates physically equivalent solutions where $T$, $J$ are changed with ${\cal T}$ fixed. We can use this symmetry to set $r_H = 1$. 

To summarize, we set $r_H=1$ and explore the one parameter family of solutions parameterized by $\phi_0 = \phi(1)$. From the form of the potential, the value of the scalar field at the horizon is allowed to vary in between the two extrema i.e. $0 < \phi_0 < \phi_{min}$ where  $\phi_{min} = \sqrt{\frac{-3m^2}{4c}}$. Taking $\phi_0 \to 0$ corresponds to the high-temperature limit, since the solution with $\phi_0 = 0$ everywhere corresponds to the UV CFT. The opposite limit $\phi_0 \to \phi_{min}$ corresponds to the low temperature limit of the field theory.

For each value of $\phi_0$, we calculate\footnote{The formula for $J$ is valid for $\Delta > 2$. For $\Delta < 2$, $J$ can still be obtained as the coefficient of $r^{\Delta - 4}$ in the expansion of $\phi(r)$ for large $r$, but this is no longer the leading term.}
\begin{equation}
\label{eq:rHone}
    T = {\frac{1}{4 \pi}} f'(1) e^{\chi_\infty/2} ,\qquad J = \lim_{r \to \infty} r^{4 - \Delta} \phi(r) ,\qquad \mathcal{T} = T J^{1/(\Delta - 4)} \; .
\end{equation}

The entropy density is obtained using the Bekenstein formula $S = A/(4 G_5)$. With $r_H = 1$, the horizon area is the same as the field theory volume, so we have simply\footnote{Recall that we are fixing $r_H=1$. To see the temperature dependence of entropy density, we should perform a conformal rescaling to fix $J$. We will instead just work with the dimensionless quantities we have defined.}
\begin{equation}
    s = \frac{1}{4G_5} \; .
\end{equation}
The dimensionless entropy is 
\begin{equation}
    {\cal S} = \frac{1}{4G_5 T^3} \; 
\end{equation}
where $T$ is taken from (\ref{eq:rHone}).
In the method we are using, $e$ and $p$ are obtained via thermodynamic relations as explained in the previous section, so it is not necessary to explicitly perform any holographic renormalization. We could alternatively have obtained $p$ by evaluating the regularized action (per unit field theory volume) after adding an appropriate set of counterterms \cite{holorg, hologren}.

\section{Results}

\begin{figure}[H]
\centering
\includegraphics[scale=0.45]{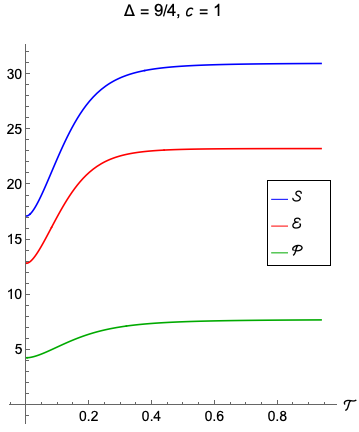}
\caption{Dimensionless quantities (rescaled by $4G_5/L_{AdS}^3$) for energy density, pressure and entropy density as a function of temperature. For large and small temperatures, $\mathcal{S} = \mathcal{E} + \mathcal{P}$ where the theory approaches a CFT.}
\label{eps}
\end{figure}

In Figure \ref{eps}, we show the behavior of ${\cal E}({\cal T})$, ${\cal P}({\cal T})$, and ${\cal S}({\cal T})$ for an example field theory with $\Delta = 9/4$, and $c=1$. 

The decrease in ${\cal S}$ (the coefficient of $T^3$ in the entropy) from high temperature to low temperature is consistent with the expected reduction in degrees of freedom in the RG flow from UV to IR. We observe the expected CFT relations $4/3 {\cal E}_{CFT} = 4 {\cal P}_{CFT} = {\cal S}_{CFT}$ in both the UV and IR. 

\begin{figure}[H]
\centering
\begin{subfigure}{0.5\textwidth}
  \centering
  \includegraphics[scale=0.45]{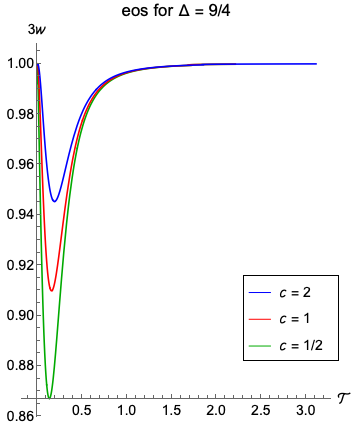}
  \caption{}
\end{subfigure}%
\begin{subfigure}{0.5\textwidth}
  \centering
  \includegraphics[scale=0.45]{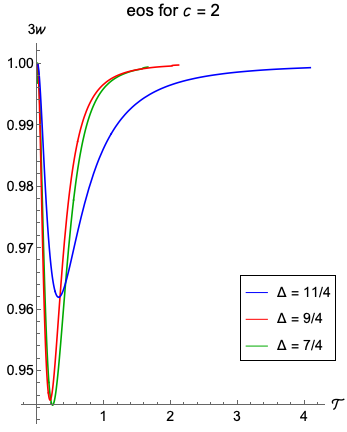}
  \caption{}
\end{subfigure}
\caption{Dependence of the equation of state parameter $w$ on dimensionless temperature for various model parameters. Both high temperature and low temperature limits correspond to conformal behaviour $w=1/3$. The intermediate behaviour depends on  the dimension $\Delta$ of the perturbing relevant operator and the coefficient $c$ of the quartic term in the bulk scalar potential.}
\label{fig:eos}
\end{figure}

The equation of state parameter $w$ is plotted against ${\cal T}$ for a variety of parameter values in Figure \ref{fig:eos}. We see that in each case, the parameter approaches a conformal value $w = 1/3$ for both high and low temperature and dips to smaller values in between. A qualitatively similar behavior was found in  \cite{Casalderrey_Solana_2021} for a different non-conformal holographic model.

The minimum value of $w$ decreases both for decreasing $\Delta$ with fixed $c$ and for decreasing $c$ with fixed $\Delta$. In each case, $V(\phi_{min})$ decreases, so roughly, we get a more significant dip in $w$ when there is a more significant reduction in the number of degrees of freedom along the RG flow.

\begin{figure}[H]
\centering
\includegraphics[scale=0.45]{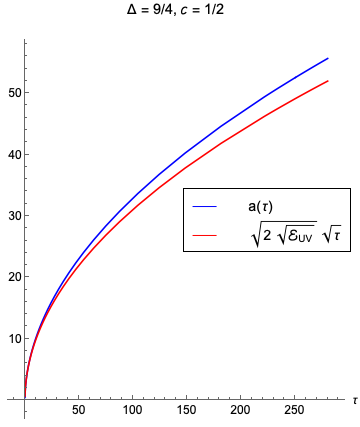}
\caption{The scale factor approaches the conformal behaviour $\sqrt{\tau}$ in the UV and for later times it starts to deviate from that.}
\label{scalefactor}
\end{figure}

To exhibit the effects of the RG flow on the evolution of the scale factor as compared to the $s^{1/2}$ evolution in a CFT, we have plotted in Figure \ref{scalefactor} the scale factor evolution in an example with $\Delta = 9/4$ and $c=1/2$, together with the scale factor evolution in the UV CFT for that theory, assuming that the early-time behavior matches.

We see that the dip in the deceleration parameter at intermediate temperature scales results in greater increase in the scale factor for the non-conformal field theory.

\subsection{Validity of the quasi-static approximation}

In our analysis, we have made use of the equilibrium equation of state. For sufficiently rapid cosmological expansion, the quantum field theory can be driven out of equilibrium. In this case, it is still possible to study the physics holographically, but this requires an analysis where the holographic quantum field theory is studied on an FRW spacetime whose geometry is fixed by a self-consistency equation that requires the Einstein equation to be satisfied with the holographically calculated stress-energy tensor. This is equivalent to studying the cosmological evolution via the 4D semiclassical Einstein equation. See \cite{2022paper, dynamicalinflaton, Koyama2001}  for examples of this approach.

For a CFT, physics on any FRW background is related to the finite-temperature physics on Minkowski space by a Weyl transformation. In the holographic setting, the Weyl transformation corresponds to a change of coordinates in the bulk, so to study the cosmological physics of the field theory  via the semiclassical approach, we only need the planar black hole solution. The result of this analysis is that the quasistatic approximation is good wherever the curvature is significantly larger than Planck scale. 

In our case, provided that the RG scale is significantly less than the Planck scale, there will be a time after the big bang when the curvature is much smaller than Planck scale where the field theory is still well-approximated by the UV CFT. During this time, the quasistatic approximation will be valid. At late times, when the quantum field theory physics is controlled by the IR CFT, the curvatures will be small so the quasistatic approximation is again valid. Thus we only need to be concerned about the intermediate times when the temperature scale is of order the RG scale. 

\begin{figure}[H]
\centering
\begin{subfigure}{.5\textwidth}
  \centering
  \includegraphics[scale=0.45]{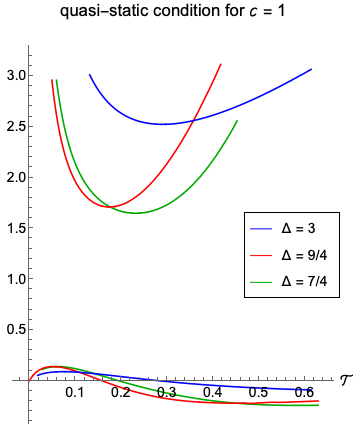}
  \caption{}
\end{subfigure}%
\begin{subfigure}{.5\textwidth}
  \centering
  \includegraphics[scale=0.45]{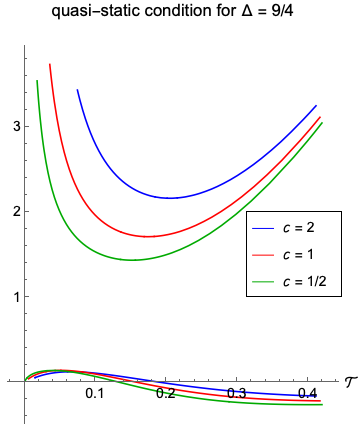}
  \caption{}
\end{subfigure}
\caption{Larger values of c and larger values of $\Delta$ admit a more accurate quasi-static description; higher curves are $\left| \dot{a}/a \right|$ and lower ones are $\left| \dot{w}/w \right|$ and it can be seen that for small and large values of $\phi_0$ where the theory is closest to being conformal the condition Eq.\ref{quasi} is satisfied.}
\label{fig:quasi}
\end{figure}

We can use the deviation of the $w$ parameter from 1/3 as a guide to when the field theory has a significant scale dependence. We expect that if the timescale $w/\dot{w}$ associated with changes in $w$ is large compared to the cosmological timescale $t_H = 1/H$, the quasistatic approximation that we have been using should be reliable. Thus, we would like to check whether  
\begin{equation}
    \left| \frac{\dot{w}}{w} \right| \ll \left|\frac{\dot{a}}{a} \right| .
    \label{quasi}
\end{equation}
As shown in Figure \ref{fig:quasi}, this condition always holds for early and late times (large and small temperatures). At intermediate times, it becomes increasingly valid for larger values of $c$ and $\Delta$ where the RG flow involves a less substantial change in the number of degrees of freedom. Since the ratio between the right side and left side of (\ref{quasi}) is always at least of order 10 for the parameter values we considered, we expect that a more exact analysis would yield only small corrections.

\subsection{Trace–anomaly effects in the quasi–static regime}

For a CFT (or a QFT that behaves as a CFT in the UV) on an FRW background, there are additional contributions to the stress-energy tensor due to the conformal anomaly. We have ignored these so far, since we expect them only to be relevant at very early times before the relevant perturbation comes into play.

In this subsection we analyze more explicitly when this is justified, finding that as long as
\begin{equation}
  J^{\frac{1}{4-\Delta}}\;\ll\;\frac{M_{\rm Pl}}{N}\;,
  \label{eq:Jbound}
\end{equation}
the transient anomaly–driven effects always die out \emph{before} the relevant
deformation becomes important.

\paragraph{Holographic stress tensor and trace anomaly.}
Using the standard AdS$_5$/CFT$_4$ dictionary 
$l^{3}/G_{5}=2N^{2}/\pi$, the renormalised boundary stress tensor on a spatially
flat FRW metric 
\(\mathrm{d}s^{2}=-\mathrm{d}t^{2}+a^{2}(t)\,\mathrm{d}\vec x^{2}\) is
\begin{align}
  T^{0}_{\;0} &=
    -\frac{3N^{2}}{32\pi^{2}}
    \frac{\dot a^{4}+E}{a^{4}},
&
  T^{i}_{\;j} &=
    \frac{N^{2}}{32\pi^{2}}
    \frac{\delta^{i}_{\;j}\,[\dot a^{2}(\dot a^{2}-4a\ddot a)+E]}{a^{4}},
  \label{eq:Tij_full}
\end{align}
where $E\ge0$ is an integration constant fixed by initial
conditions.  
Taking the trace of~\eqref{eq:Tij_full} reproduces the CFT$_4$ anomaly
\cite{Henningson_1998, Balasubramanian_1999}
\begin{equation}
  \langle T^{\mu}_{\;\mu}\rangle
  = -\frac{N^{2}}{\pi^{2}}\!\left(E_{4}+I_{4}\right),
  \label{eq:anomaly}
\end{equation}
with
\(
E_{4}=R_{\mu\nu\rho\sigma}R^{\mu\nu\rho\sigma}-4
            R_{\mu\nu}R^{\mu\nu}+R^{2}
\)
(the four–dimensional Euler density) and
\(
I_{4}=C_{\mu\nu\rho\sigma}C^{\mu\nu\rho\sigma}
\)
(the square of the Weyl tensor).
For a flat FRW background $I_{4}=0$, and~\eqref{eq:anomaly} reduces to
\(
\langle T^{\mu}_{\;\mu}\rangle=
 -\frac{3N^{2}}{16\pi^{2}}\,H^{2}(\dot H+H^{2}),
\)
where \(H=\dot a/a\).

\paragraph{Einstein equation and exact time dependence.}
Coupling~\eqref{eq:Tij_full} to four-dimensional gravity via
$G^{0}_{\;0}=8\pi G_{N}T^{0}_{\;0}$ gives
\begin{equation}
  a^{2}=\frac{N^{2}G_{N}}{4\pi}\Bigl(\dot a^{2}+E/\dot a^{2}\Bigr).
  \label{eq:Friedmann_exact}
\end{equation}
Solutions with different values of $E$ are related to each other by a rescaling \(a\to E^{1/4}a\)), so we will choose a value 
\(E_{\star}=4\pi^{2}/(G_{N}^{2}N^{4})\;\). The equation (\ref{eq:Friedmann_exact}) has two branches of solutions, an inflating branch and one associated with ordinary late-time cosmological evolution. We are interested in the latter, which is governed by
\begin{equation}
  \dot a^{2}
  =\frac{2\pi}{G_{N}N^{2}}
   \Bigl(a^{2}-\sqrt{a^{4}-1}\Bigr).
  \label{eq:a_dot_solution}
\end{equation}

\paragraph{Early and late times.}

A characteristic feature of the solution is that there is some initial time $t_0$ with finite energy density for which the solution cannot be extended to earlier times.

Our choice of $E_*$ corresponds to taking $a = 1$ at this initial time. We have energy density
$T^{0}_{\;0}(t_{0})=-3/(4N^{2}G_{N}^{2})$ here; this is the maximum energy density
along the trajectory.  For $a\gg1$ the solution expands as
\[
\dot a^{2}\simeq\frac{\pi}{G_{N}N^{2}\,a^{2}}
\quad\Longrightarrow\quad
a^{2}(t)\simeq2\sqrt{\frac{\pi}{G_{N}N^{2}}}\;t,
\]
so the universe asymptotes to the standard radiation law $a\propto t^{1/2}$.
\paragraph{Decay of the anomaly contribution.}

Splitting \(T^{0}_{\;0}\) into an ``anomaly’’ part
\(\rho_{A}\propto\dot a^{4}/a^{4}\) and a ``radiation’’ part
\(\rho_{R}\propto E/a^{4}\), equation~\eqref{eq:a_dot_solution} implies
\[
\frac{\rho_{A}}{\rho_{R}}
 =\frac{\dot a^{4}}{E}
 =\Bigl(a^{2}-\sqrt{a^{4}-1}\Bigr)^2
 \xrightarrow{a\gg1}\frac{1}{4\,a^{4}}.
\]
Hence the anomaly contribution is approximately equal to the radiation term at the initial time and is suppressed by an extra factor $a^{-4}$ at later times.  Taking the equality point as a marker, the energy density at the end of the anomaly era
is
\begin{equation}
  \rho_{\text{end}}
  \;\simeq\;  \epsilon \frac{1}{G_{N}^{2}N^{2}},
  \label{eq:rho_end}
\end{equation}
for some small $\epsilon$.

\paragraph{Relevant deformation versus anomaly.}
For a relevant deformation \(J\mathcal O\) with ${\cal O}$ of dimension \(\Delta<4\)
we have an associated energy scale
\(\rho_{J}\sim N^{2}J^{\,4/(4-\Delta)}\).
To ensure that the RG flow becomes important only \emph{after} the
trace–anomaly can be ignored, we require
\[
\rho_{J}\;<;\rho_{\text{end}}
          \;\simeq\;\epsilon \frac{1}{G_{N}^{2}N^{2}}
\quad\Longrightarrow\quad
J^{\frac{1}{4-\Delta}}\;\ll\;\frac{M_{\text{pl}}}{N}.
\label{eq:Jbound_main}
\]
Thus, provided that the energy scale associated with relevant perturbation is below the Planck scale by at least a factor of $N$, we can ignore the effects of the conformal anomaly for our analysis.

\section{Discussion}

Using the framework of AdS/CFT, we have characterized the effects on cosmological evolution of a strongly coupled non-conformal quantum field theory whose RG flow interpolates between UV and IR CFTs. The analysis showed a distinctive signature: the equation–of–state parameter approaches the conformal value $w=1/3$ at very high and very low temperatures, but dips to smaller values when the temperature is comparable to the scale set by the source for the relevant operator. 

The temporary dip has a simple physical origin. As the system cools to the scale associated with the relevant perturbation, some of the degrees of freedom of the CFT begin to behave more like massive non-relativistic matter with $w=0$ before dropping out completely in the IR. 
Once the temperature falls far enough that the theory is governed by the IR fixed point, the remaining excitations are again effectively massless, radiation dominates, and $w$ returns to $1/3$.  
The depth of the dip is governed by how drastically the number of active relativistic degrees of freedom is reduced along the RG flow, and this is captured by the change of the bulk potential $V(\phi_{\min})$.  For fixed $c$, decreasing the scaling dimension $\Delta$ makes the deformation more relevant and therefore eliminates a larger fraction of UV degrees of freedom, pushing the minimum of $w$ lower.  For fixed $\Delta$, decreasing $c$ has a similar effect: it deepens the potential and therefore gives a larger fraction of the degrees of freedom that are removed during the RG-flow between the UV and IR theories, again producing a more pronounced reduction in $w$ in the intermediate stages.

While there is no particular reason to expect that such a strongly coupled non-conformal theory exists in the dark sector, it is useful to keep this possibility in mind since it gives a behavior more general than the standard matter, radiation or cosmological constant.  

Our analysis was restricted to the quasi-static regime in which equilibrium physics is assumed. We have provided evidence that this approximation is reasonable for the theories we have considered. We also examined the effects of the trace anomaly and argued that we can ignore this during the period of evolution where the RG-flow is relevant provided that $J^{\frac{1}{4-\Delta}}\ll \frac{M_{\text{pl}}}{N}$. In this case, the conformal anomaly only affects the very early evolution where the energy density is high enough that the relevant perturbation is unimportant.

For more accuracy, or in cases where (perhaps because of other matter components) the cosmological evolution becomes fast relative to the time scale associated with the field theory, methods similar to \cite{2022paper, dynamicalinflaton, Koyama2001} could be used to study the evolution using the full semiclassical Einstein equation.

\newpage
\bibliographystyle{jhep}
\newpage
\bibliography{references}

\end{document}